\UseRawInputEncoding
\documentclass[aps,prl,reprint,noaffiliation,groupedaddress,amsmath,amssymb,]{revtex4-2}
\usepackage{graphicx}% 
\usepackage{float}
\usepackage[x11names]{xcolor}
\usepackage[colorlinks,citecolor=blue]{hyperref}

\begin{document}
	
	\title{Spin-Orbit-Locking Chiral Bound States in the Continuum}

	\author{Xingqi Zhao$^{1}$}
	\author{Jiajun Wang$^{1}$}
	\email{jiajunwang@fudan.edu.cn}
	\author{Wenzhe Liu$^{2,3}$}
	\author{Zhiyuan Che$^{1}$}
	\author{Xinhao Wang$^{1}$}
	\author{C.T.Chan$^{2}$}
	\author{Lei Shi$^{1,3,4,5}$}
	\email{lshi@fudan.edu.cn}
	\author{Jian Zi$^{1,3,4,5}$}
	\email{jzi@fudan.edu.cn}
	%\thanks{}
	\affiliation{$^{1}$State Key Laboratory of Surface Physics, Key Laboratory of Micro- and Nano-Photonic Structures
		(Ministry of Education) and Department of Physics, Fudan
		University, Shanghai 200433, China}
	\affiliation{$^{2}$Department of Physics, The Hong Kong University of Science and Technology, Hong Kong 999077, China}
	\affiliation{$^{3}$Institute for Nanoelectronic devices and Quantum computing, Fudan University, Shanghai 200438, China}
	\affiliation{$^{4}$Collaborative Innovation Center of Advanced Microstructures, Nanjing University, Nanjing 210093, China}
	\affiliation{$^{5}$Shanghai Research Center for Quantum Sciences, Shanghai 201315, China}
	
	%\date{\today}
	
	\begin{abstract}
	{Bound states in the continuum (BICs), which are confined optical modes exhibiting infinite quality factors and carrying topological polarization configurations in momentum space, have recently sparked significant interest across both fundamental and applied physics.} Here we show that breaking time-reversal symmetry by external magnetic field enables a new form of chiral BICs with spin-orbit locking. Applying a magnetic field to a magneto-optical photonic crystal slab lifts doubly degenerate BICs into a pair of chiral BICs carrying opposite pseudo-spins and orbital angular momenta. Multipole analysis verifies the non-zero angular momenta and reveals the spin-orbital-locking behaviors. In momentum space, we observe ultrahigh quality factors and near-circular polarization surrounding chiral BICs, enabling potential applications in spin-selective nanophotonics. Compared to conventional BICs, the magnetically-induced chiral BICs revealed here exhibit distinct properties and origins, significantly advancing the topological photonics of BICs by incorporating broken time-reversal symmetry.
	\end{abstract}
	
	\maketitle
	
	Optical bound states in the continuum (BICs) have recently gained significant attention due to both their fascinating underlying physics and potential applications \cite{hsu2016bound,kang2023applications}. Initially recognized for their ability to confine light, led by their infinite quality \mbox{(Q-)} factors \cite{hsu2013observation,jin2019topologically}, BICs have also been discovered to be topological defects in momentum space, carrying topological charges as determined by the winding number of surrounding polarization states \cite{zhen2014topological,zhang2018observation,doeleman2018experimental}. The exceptionally high Q-factors and topological polarization configurations enabled by BICs have spurred progress in numerous research domains: sensing \cite{tittl2018imaging}, lasing \cite{kodigala2017lasing,huang2020ultrafast,rong2023spin}, Bose-Einstein condensation \cite{ardizzone2022polariton,Gianfrate2024Reconfigurable}, and spin-orbit interactions of light \cite{wang2020generating,wang2022spin}, etc. 
	This increasing utilization of BICs emphasizes the necessity for their effective manipulation, which in turn leads to the unveiling of novel physical effects. Researchers have embarked on the journey of modulating and evolving BICs, primarily through adjusting structural parameters and disrupting point group symmetry \cite{gomis2017anisotropy,koshelev2018asymmetric,liu2019circularly,liu2019highQ,yoda2020generation,ye2020singular,kang2021merging,zeng2021dynamics}. This has resulted in the emergence of optical modes with unique properties, such as chiral quasi-BICs or BICs \cite{gorkunov2020metasurfaces,overvig2021chiral,zhang2022chiral,shi2022planar,chen2023observation} and unidirectional guided resonances \cite{yin2020observation,yin2023topologicalugr}.
	
	Yet, an intriguing aspect that remains to be thoroughly investigated is the influence of time-reversal ($\mathcal{T}$) symmetry breaking on BICs, especially under the condition of magnetic field. In topological photonics, {breaking $\mathcal{T}$ symmetry is known to result in the emergence of spin-dependent topological properties like spin-momentum locking edge states and spin-orbit-locking emissions \cite{wang2009observation,Klembt2018exciton,CarlonZambon2019optical,ozawa2019topological}}. Understanding how this $\mathcal{T}$ symmetry breaking with external magnetic field affects the intrinsic topological properties of photonic crystal slabs and how BICs respond to these changes is crucial. 
	
	\begin{figure*}[htbp]
		\centering
		\includegraphics{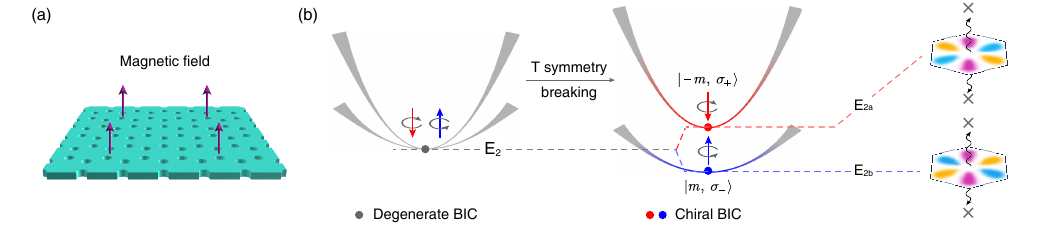}% Here is how to import EPS art
		\caption{(a) A MO PhC slab consisting of a honey-comb lattice of air holes. An external magnetic field is applied perpendicular to the slab. (b) The evolution from doubly degenerate BICs (gray point) to paired chiral BICs (red and blue points) by $\mathcal{T}$ symmetry breaking. Without the magnetic field, the $\mathrm{E_2}$ representation corresponds to doubly degenerate BICs. When the $\mathcal{T}$ symmetry is broken by the external magnetic field, double degeneracy is lifted to form a pair of chiral BICs, corresponding to $\mathrm{E_{2a}}$ and $\mathrm{E_{2b}}$ representations.}
		\label{Fig1}
	\end{figure*}
	
In this work, we are focusing on the doubly degenerate BICs in photonic crystal (PhC) slabs made of magneto-optical (MO) materials. {We reveal that, when an external magnetic field is applied to the MO PhC slab, the Zeeman-like splitting lifts doubly degenerate BICs into a pair of spin-orbit-locking chiral BICs. These chiral BICs carry both pseudo-spins and orbital angular momenta, with a specific pseudo-spin being locked to a nonzero orbital angular momentum. The pseudo-spins correspond to circular polarizations and the orbital angular momenta correspond to the electromagnetic multipolar components of chiral BICs.} These spin-orbit-locking chiral BICs exhibit evident differences from conventional BICs in both physical origins and intrinsic properties, prompting further exploration of in  topological effects of light when involving $\mathcal{T}$ symmetry.
	
The original spin degeneracy connected by $\mathcal{T}$ symmetry usually stands one of the cornerstones of topological physics, from which novel spin-related topological phenomena emerge after lifting the degeneracy by breaking $\mathcal{T}$ symmetry \cite{ozawa2019topological}. {For open boundary systems, $\mathcal{T}$ symmetry breaking and degeneracy lifting can bring novel phenomena for BICs as well. In a honeycomb PhC slab shown in Fig. \ref{Fig1}(a), there are doubly degenerate BICs guaranteed by $C_{6v}\&\mathcal{T}$ symmetry, corresponding to the $\mathrm{E_2}$ representation \cite{sakoda2005optical,doiron2022realizing}. When the PhC slab is composed of MO material, applying an external magnetic field perpendicular to the slab will break the $\mathcal{T}$ symmetry (as well as the $\sigma_v$ symmetry), while the $C_6$ symmetry remains preserved.} In analogy to the Zeeman splitting in electronic systems, the  external magnetic field here will lift the degeneracy of the MO PhC slab. The doubly degenerate modes will be split to form a pair of chiral modes with opposite pseudo-spins, which are typically characterized by two opposite circular polarizations, i.e., right-handed and left-handed circular polarized (RCP and LCP) waves. In this case, with the absence of $\mathcal{T}$ symmetry, the previous $\mathrm{E_2}$ representation will split into a pair of one-dimensional representations $\mathrm{E_{2a}}$ and $\mathrm{E_{2b}}$ \cite{dresselhaus2007group}, corresponding to the paired single modes. The spatial symmetries of the two modes still mismatch with any plane wave in free space (detailed symmetry analysis is provided in the Supplementary Material \cite{supplemental}). At $\mathrm{\Gamma}$ point, the lifted single modes are symmetry-protected chiral BICs. When slightly off the $\mathrm{\Gamma}$ point, the modes become radiative and then their pseudo-spins can be represented as LCP and RCP waves in free space.
	
Moreover, the magnetically induced  $\mathcal{T}$ symmetry breaking in this MO PhC can give rise to a scenario of spin-orbit locking. Here, the orbital angular momenta of these chiral BICs, manifested by their azimuthally accumulating phase profiles, can be revealed by the group representations. For the separated modes of $\mathrm{E_{2a}}$ and $\mathrm{E_{2b}}$ representation, their phase has opposite winding configurations (the right panel of Fig. \ref{Fig1}(b)) around the rotation axis, which can be viewed to correspond to opposite orbital angular momenta. To sum up, {the magnetically induced $\mathcal{T}$ symmetry breaking in the MO Phc slab } would lead to spin-split bands with lifted degeneracy and a pair of spin-orbit-locking chiral BICs, which carry both opposite pseudo-spins and orbital angular momenta. {Beside the symmetry analysis, we also use an effective Hamiltonian model to give a clear physical picture for the generation mechanism of chiral BICs. The external magnetic field will induce coupling between modes on two bands, resulting in the lifted degeneracy and the spin-orbit-locking behaviors. Details are in the Supplementary Material \cite{supplemental}.}
	
\begin{figure}[h]
		\includegraphics{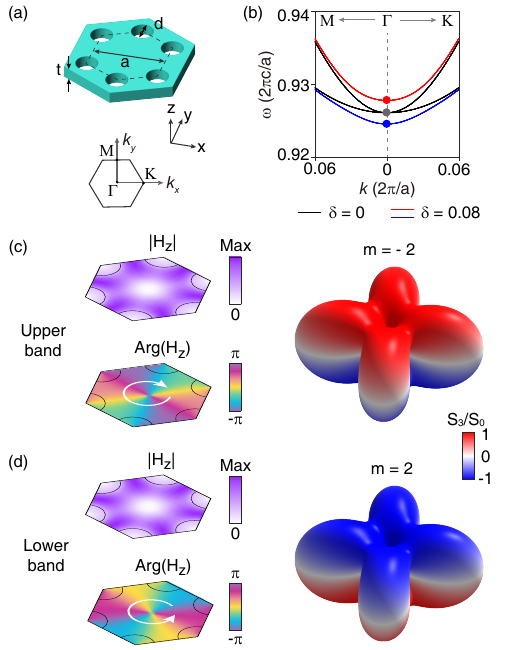}% Here is how to import EPS art
		\caption{(a) A MO PhC slab with $C_6$ symmetry in an external magnetic field applied perpendicular to the slab. (b) The evolution from doubly degenerate BICs to paired chiral BICs with the application of external magnetic field. (c), (d) In-plane field distributions and corresponding multipoles of the (c) upper band and (d) lower band.}
		\label{Fig2}
	\end{figure}

	We then give a specific example to demonstrate the proposed physical picture. As shown in Fig. \ref{Fig2}(a), the lattice constant of the MO PhC slab is denoted as $a$, and the diameter of air holes $d$ and the thickness of the slab $t$ are set to $0.333 a$ and $0.190 a$, respectively. In the simulation, $a$ is set to be $840$ $\mathrm{nm}$. With a magnetic field applied perpendicular to the slab (along the $z$ direction), the relative dielectric constant of the MO material takes the following form \cite{buschow2003handbook,haider2017review}:
	\begin{equation}
		\overset{\leftrightarrow}{\boldsymbol{\varepsilon}_r}=\begin{pmatrix}
			\varepsilon & -\mathrm{i}\delta & 0\\
			\mathrm{i}\delta & \varepsilon & 0\\
			0 & 0 & \varepsilon\\
		\end{pmatrix}.
	\end{equation}
	Here, $\varepsilon$ is set to be 4 in our calculations, and $\delta$ can be dynamically controlled by the strength of the magnetic field. The relative permeability $\mu_r$ is set to be $1$. Without the magnetic field, i.e., $\delta=0$, doubly degenerate modes corresponding to $\mathrm{E_2}$ representation can be found in this PhC slab. In Fig. \ref{Fig2}(b), the black lines show corresponding band dispersions in this structure without external magnetic field {(gray point denotes the degenerate BICs)}. By applying the external magnetic field along the $z$ direction, the off-diagonal term of the dielectric constant will have a non-zero value. For example, with $\delta=0.08$, the previously degenerate bands will be lifted, as illustrated in Fig. \ref{Fig2}(b), where the red and blue lines correspond to the dispersions of the upper and lower bands. At the $\mathrm{\Gamma}$ point, previous degenerate BICs are then spilt to be a pair of chiral BICs (marked by red and blue points). Note that, we only show the real part of the eigen-frequency here. The imaginary part and detailed evolution of the band dispersion with the increasing of $\delta$ are provided in the Supplementary Material \cite{supplemental}.

      The field profiles of the paired chiral BICs in the $\sigma_z$ mirror plane are shown in Fig. \ref{Fig2}(c) and (d). For the at-$\mathrm{\Gamma}$ chiral BIC on the upper band, it corresponds to the $\mathrm{E_{2a}}$ representation of $C_6$ group. In the left panel of Fig. \ref{Fig2}(c), we display the in-plane distributions of both the absolute value and phase of $H_z$. When rotating counterclockwisely around the $z$ axis by $60$ degrees, the absolute value of $H_z$ is invariant, while the phase of $\mathrm{E_{2a}}$ representation will change by $-2\pi/3$ (see the Supplementary Material for details \cite{supplemental}), resulting in a total phase winding of $-4\pi$. In contrast, for the at-$\mathrm{\Gamma}$ chiral BIC on the lower band, the total phase winding in the counterclockwise direction is $4\pi$. The opposite near-field phase windings around the $z$ axis have preliminarily revealed that there are orbital angular momentum components in these chiral BICs. 
	
	Furthermore, we employ the multipolar expansion to analyze the {underlying physics} of chiral BICs. Their whole field profiles can be expanded as multipoles characterized by vector spherical harmonics $\mathbf{N}_{lm}$ and $\mathbf{M}_{lm}$ \cite{chen2019singularities,Sadrieva2019Multipolar,chen2020line} (corresponding to electric and magnetic multipoles respectively):
	\begin{equation}
		\begin{aligned}
			\mathbf{N}_{lm}&=[\tau_{lm}(\theta)\hat{\boldsymbol{e}}_\theta +\mathrm{i} \pi_{lm}(\theta)\hat{\boldsymbol{e}}_\varphi] \frac{[krz_l(kr)]^\prime}{kr}\mathrm{e}^{\mathrm{i}m\varphi}\\
			\mathbf{M}_{lm}&=[\mathrm{i} \pi_{lm}(\theta)\hat{\boldsymbol{e}}_\theta - \tau_{lm}(\theta)\hat{\boldsymbol{e}}_\varphi] z_l(kr)\mathrm{e}^{\mathrm{i}m\varphi}\\
		\end{aligned},
	\end{equation}
	where $\tau_{lm}(\theta)=\frac{\mathrm{d}}{\mathrm{d\theta}}P_l^m(\sin{\theta})$, $\pi_{lm}(\theta)=\frac{m}{\sin{\theta}}P_l^m(\sin{\theta})$ ($P_l^m$ is the associated Legendre
	polynomials), and  $z_l$ is the spherical Bessel function or Hankel function \cite{chen2019singularities,chen2020line}. A detailed explanation is provided in the Supplementary Material \cite{supplemental}. It could be seen the orbital angular momentum is inherent in the formalism of spherical harmonics, where $m$ corresponds to the effective magnetic quantum number \cite{tischler2012role}. For the chiral BIC on the upper band (the right panel of Fig. \ref{Fig2}(c)), $\mathbf{N}_{2,-2}$ is the major dominant multipole, whose effective magnetic quantum number is $-2$. While the major dominant multipole is $\mathbf{N}_{2,2}$ for chiral BIC on the lower band, carrying opposite effective magnetic quantum number of $2$. {The multipole components offer deep insights into the radiation properties of chiral BICs and modes around the $\mathrm{\Gamma}$ point.} For both chiral BICs, the radiation of their multipoles vanish along the $z$ axis, confirming their nature as radiation singularities. Considering radiation parts of the multipoles, the corresponding spin properties of light (characterized by the normalized third Stokes parameter $S_3/S_0$ defined along the radiation direction) are marked by the red and blue color map. When the direction slightly deviates from the $z$ axis, we observe high degree of circular polarization in radiation. And the spin components for this pair of chiral BICs are also opposite to each other. We not only confirm the spin-orbit locking in chiral BICs based on the multipole analysis, but also expect chiral characteristics in the radiation.
	
	\begin{figure}[b]
		\includegraphics{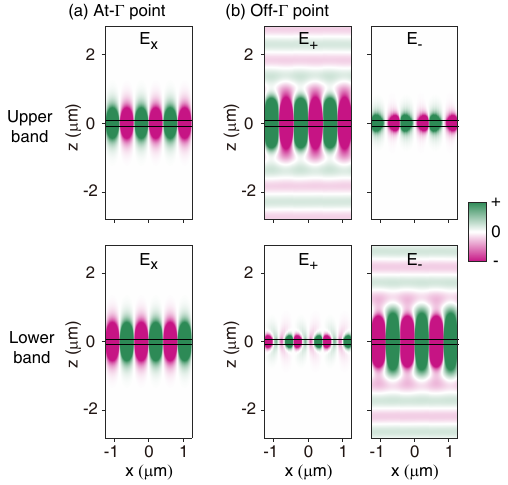}% Here is how to import EPS art
		\caption{ (a) Side views of electric field profiles ($x$ component) of chiral BICs on the upper band (upper panel) and lower band (lower panel). (b) Side views of electric field profiles (circularly polarized components) of nearby leaky modes off the $\mathrm{\Gamma}$ point. The $\delta$ is set to be 0.08 and the exhibited side views contain three unit cells.}
		\label{Fig3}
	\end{figure}
	
	For chiral BICs at the $\mathrm{\Gamma}$ point, the $x$ components of electric field profiles are exhibited in Fig. \ref{Fig3}(a). The results of $y$ components of electric field profiles are shown in the Supplementary Material \cite{supplemental}. The electric fields are confined in the near field and there is no outgoing wave, confirming the non-radiative nature of chiral BICs. For modes slightly off the $\mathrm{\Gamma}$ point, e.g., {with wavevector $(k_x,k_y )=(0.01,0)\times2\pi/a$ (see Fig. \ref{Fig3}(b))}, they become radiative.
	To exhibit the chiral properties of radiation waves in the far field, here, we decompose the electric field into two orthogonal parts:
	\begin{equation}
		E_{\pm}=\langle \hat{\boldsymbol{e}}_x\pm \mathrm{i} \hat{\boldsymbol{e}}_y| {\boldsymbol{E}} \rangle,
	\end{equation}
	where $\hat{\boldsymbol{e}}_x$ ($\hat{\boldsymbol{e}}_y$) is the basis vector of the $x$ ($y$) direction, and $E_+$ ($E_-$) corresponds to RCP (LCP) components defined in the $x-y$ plane. For the leaky mode near the chiral BIC on the upper band, the $E_+$ part is dominant in the far-field radiation waves while the $E_-$ part vanishes. In contrast, for the leaky mode near the chiral BIC on the lower band, the $E_-$ part is dominant in the far-field radiation waves while the $E_+$ part vanishes. The radiation waves of leaky modes near chiral BICs are nearly circularly polarized, with opposite circular polarizations for the upper band and lower band. These results agree well with the expectations from the multipole results in Fig. \ref{Fig2}(c-d).
	
	BICs are also known as topological defects in momentum space, carrying topological charges. The topological charge $q$ describes the winding number of polarization axis of leaky modes around BICs, which is defined as \cite{zhen2014topological}:
	\begin{equation}
		q=\frac{1}{2\pi}\oint_{L}\mathrm{d}\boldsymbol{k}_{\parallel}\cdot\nabla_{\boldsymbol{k}_{\parallel}}\phi(\boldsymbol{k}_{\parallel}).
	\end{equation}
	Here, $L$ is a close loop around the BIC, and $\phi(\boldsymbol{k}_{\parallel})$ is the azimuthal angle of the polarization states' major axis. With applied external magnetic field in the MO PhC slab, there are also novel evolutions for winding distributions around BICs in momentum space.
	
	\begin{figure}[t]
		\includegraphics{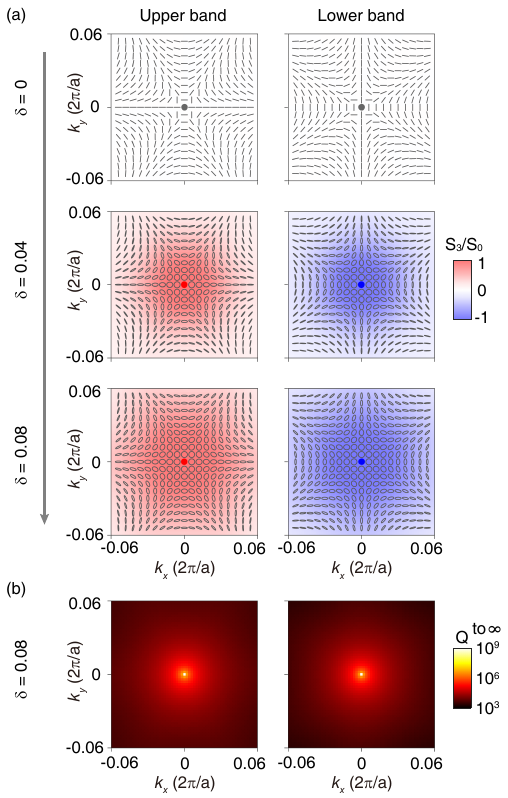}% Here is how to import EPS art
	    \caption{(a) Evolutions of the polarization distribution in momentum space of the upward radiations with the enhancing of magnetic field. The degree of circular polarization of mode is represented by the background color. (b) The Q-factor distribution in momentum space with $\delta=0.08$.}
	\label{Fig4}
	\end{figure}
	
	Fig. \ref{Fig4}(a) shows the evolutions of polarization distributions in momentum space with the increasing of the applied external magnetic field. The first row of Fig. \ref{Fig4}(a) shows the case without the external magnetic field ($\delta=0$), where BICs are degenerate at the $\mathrm{\Gamma}$ point. The color map shows the normalized third Stokes parameter ($S_3/S_0$), characterizing the chiral properties in momentum space. It can be seen that degenerate BICs carry topological charge of $-2$. The polarization states around degenerate BICs are nearly linearly polarized, similar to conventional nondegenerate BICs. With an external magnetic field, in middle and bottom rows of Fig. \ref{Fig4}(a), the polarization states around chiral BICs show high degree of circular polarization. The increasing of external magnetic field (i.e., the increasing of $\delta$) enlarges the momentum-space distribution with high degree of circular polarization, indicating the enhanced magnetically induced coupling. In this process, the topological charge of each chiral BIC maintains $-2$, governed by the conservation law of topological charges. As approaching to the central chiral BICs, the polarization states approach circular polarization, and the winding azimuthal angles turn into a winding phase. This reflects the spin and orbit properties of chiral BICs and their locking behaviour. 
    In Fig. \ref{Fig4}(b), we also exhibit the distributions of Q-factors in momentum space with $\delta=0.08$. The Q-factors diverge at $\mathrm{\Gamma}$ point due to the chiral BICs. From the momentum-space results, we can see the chiral BICs give rise to the combination of high Q-factors and high degree of circular polarization. 

    {With peculiar properties, magnetically induced chiral BICs can inspire further explorations and applications. The inherent ultrahigh Q-factors and the spin-orbit-locking properties make the chiral BICs suitable for developing novel vortex lasers. In comparison to vector lasing via conventional BICs \cite{kodigala2017lasing,huang2020ultrafast,rong2023spin}, vortex lasing via chiral BICs in this work would have circular polarization and carry orbital angular momentum. The chiral BICs in PhC slabs also provide a nonlocal approach for achieving spin-orbit-locking lasing \cite{CarlonZambon2019optical,zhang2020tunable,Zhang2022spin}. Besides, when combined with the magnetically manipulable mechanism, chiral BICs can also bring explorations in various unique resonant effects, such as asymmetric spin-dependent beam shifts, anomalous magneto-optical Kerr effect, ultra-narrow spin filter, and magnetically controllable switcher. Some examples of these effects are given in the Supplementary Material \cite{supplemental}. Certainly, more hidden effects and applications are expected in future research.} 
    
    In conclusion, we have revealed and verified a new form of spin-orbit-locking chiral BICs enabled by magnetically induced $\mathcal{T}$ symmetry breaking in a MO PhC slab. The chiral BICs exhibit opposite pseudo-spins and angular momenta, verified through multipole analysis. The modes around BICs process both ultrahigh Q-factors and chiral far field radiations. This property is practicable for applications such as chiral emission and sensing, and will also promote potential investigations in various spin-dependent photonic researches. The orbital angular momenta of these chiral BICs can provide more possibilities for optical manipulation. Our work paves the way for 
    {exploring novel spin-related effects and applications}
    in BICs and broadening new research directions in topological photonics and spin-photonics with $\mathcal{T}$ symmetry breaking conditions.
	
	\bigskip
	
	\begin{acknowledgments}
		We thank Prof. Liu and J. Y. Wu for helpful discussions.
		This work is supported by National Key R\&D Program of China (2023YFA1406900 and 2022YFA1404800); 
		National Natural Science Foundation of China (No. 12234007, No. 12321161645, and No. 12221004); 
		Major Program of National Natural Science Foundation of China (Grant No. T2394480, T2394481);
		Science and Technology Commission of Shanghai Municipality (22142200400, 21DZ1101500, 2019SHZDZX01 and 23DZ2260100);
		Research Grants Council (RGC) of Hong Kong (CRS\_HKUST601/23).
		J. W. is further supported by China National Postdoctoral Program for Innovative Talents (BX20230079) and China Postdoctoral Science Foundation (2023M740721). 
		
		\bigskip
		
		X.Z., J.W., and W.L. contributed equally to this work.
		
	\end{acknowledgments}

\end{document}